\documentstyle[aps,multicol,epsf,epsfig]{revtex}

\def\CC{{\rm\kern.24em \vrule width.04em height1.46ex depth-.07ex
\kern-.30em C}}
\def\P{{\rm I\kern-.25em P}}
\def\RR{{\rm
         \vrule width.04em height1.58ex depth-.0ex
         \kern-.04em R}}

\def\RR{{\rm\kern.24em \vrule width.04em height1.46ex depth-.07ex
\kern-.30em R}}
\def\P{{\rm I\kern-.25em P}}
\def\RR{{\rm
         \vrule width.04em height1.58ex depth-.0ex
         \kern-.04em R}}

\newcommand{\be}{\begin{equation}}
\newcommand{\ee}{\end{equation}}
\newcommand{\bq}{\begin{eqnarray}}
\newcommand{\eq}{\end{eqnarray}}

\newcommand{\Sp}{\,\,\,\,\,\,}
\newcommand{\no}{\nonumber\\}

\newcommand{\si}{\sigma}

\newcommand{\lan}{\langle}
\newcommand{\ran}{\rangle}

\date{\today}

\begin{document}
\draft
\title{Quantum computation with trapped ions in an optical cavity}
\author{Jiannis Pachos\footnote{jip@mpq.mpg.de} and Herbert Walther}
\address{ 
Max-Planck-Institut f\"ur Quantenoptik, D-85748 Garching, Germany
}

\maketitle
\begin{abstract}
Two-qubit logical gates are proposed on the basis of two atoms trapped in a
cavity setup. Losses in the interaction by spontaneous transitions are
efficiently suppressed by employing adiabatic transitions and the Zeno
effect. Dynamical and
geometrical conditional phase gates are suggested. This method provides
fidelity and a success rate of its gates very close to unity. Hence, it is
suitable for performing quantum computation.
\end{abstract}

\begin{multicols}{2}


\vspace{0.2cm}

One of the main obstacles in realizing a quantum computer (QC) is decoherence
resulting from the coupling of the system with the environment. There are
theoretical proposals for models which avoid decoherence
\cite{pellizzari,dfs,molmer,beige}. For this purpose decoherence-free
subspaces (DFS) have been proposed in the literature for performing QC
\cite{Duan1,Zanardi,Bacon}. While they are easy to construct in the case of a
single qubit, they are more complicated for the case of an externally
controlled multipartite system. Their main decoherence channel is the ``bus''
that couples the different subsystems and is usually strongly perturbed by the
environment. In the case of an ion trap the bus is the common vibrational mode
which is subject to continuous heating. In the case of cavity QED the bus is a
cavity mode which may leak to the environment. Additionally the cavity
couples to an excited state of the atom that shows spontaneous emission. To
avoid these phenomenon it is most convenient to transfer population by
virtually populating the bus \cite{molmer,beige,plenio,haroche,solano}. Here
we present a model with atoms in an optical cavity that bypasses the
decoherence problem with, in principle, arbitrarily large fidelity and success
rate.
\begin{center}
\begin{figure}[ht]
\centerline{
 \epsffile{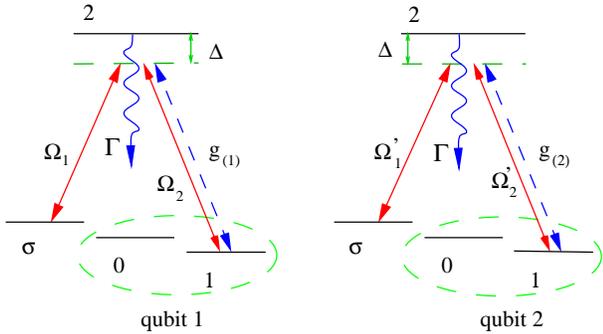}
}
\vspace{0.3cm}
 \caption[contour]{\label{at2}
Atomic levels of the two atoms and laser and cavity couplings with their
detunings. The qubits $1$, $2$ and the auxiliary states $|\sigma\ran$ are
depicted. 
}
\end{figure}
\end{center}
\vspace{-1cm}

The system presented here consists of two four-level atoms fixed inside an
optical cavity, Fig. \ref{at2}. This can be achieved by, for example, having
trapped ions in a cavity with its axis perpendicular to the ionic chain. It
is assumed that the atoms have the lower states, $|0\ran$, $|1\ran$ and
$|\sigma\ran$, which could be represented by different hyperfine levels or
Zeeman levels, and an excited state $|2\ran$ coupled individually to each
ground state by laser radiation with different polarizations or
frequencies. The atoms interact with each other via the common cavity
radiation field.

Our goal is to perform QC in such a way that coherent evolution can be
performed even for high loss rates, $\kappa$, of the cavity and relatively
large decay rates, $\Gamma$, of the excited atomic states. This is achieved
by employing an adiabatic procedure that keeps the cavity empty and the
excited state of the atoms depopulated. Information is transfered by virtual
population of these decohering atomic and cavity states. The entangling
adiabatic transfer of population between ground state levels occurs by slowly
varying the Rabi frequencies of the lasers in a counterintuitive temporal
sequence, similarly to the well-known STIRAP process for $\Lambda$ systems,
but now it is performed in the space spanned by the tensor product states of
the two atoms. Its effect is to avoid spontaneous emissions from the atoms by
adiabatically eliminating the excited levels. Even though the duration of the
two-qubit gate could be increased, the decoherence rate of the qubit states
{\it even during gate performance} is greatly suppressed.

Consider the atomic ground states $|0\rangle$ and $|1\rangle$, which span the
computational space, and the excited $|2\rangle$ as well as the Fock states
of the cavity denoted by $|n\rangle$ with $n=0,1,...\,$. For an empty cavity
tuned along the $1$-$2$ transition with equal atom-cavity coupling
$g_{(1)}=g_{(2)}=g$ \cite{comment} for both atoms the following states span a
decoherence-free subspace DFS$_c$ \cite{beige} with respect to cavity emissions:
$|00\rangle$, $|01\rangle$, $|10\rangle$, $|11\rangle$ and
$|\alpha\rangle=\big( |12\rangle -|21\rangle \big)/\sqrt{2}$. These states are
annihilated by the atom-cavity interaction Hamiltonian and hence do not
populate the cavity. Still the maximally entangling state $|\alpha
\rangle$, when populated, may 
result in atomic emission as it occupies the excited level $2$. 
Hence, the decoherence free subspace DFS$_{ca}$ with respect to both
atomic and cavity emissions is spanned only by the states
$|00\rangle$, $|01\rangle$, $|10\rangle$ and $|11\rangle$ which are
all ground states. First, we
shall review the mechanism for suppressing the effect of $\kappa$ during the
performance of a gate and then we shall show how to suppress the effect of
$\Gamma$.

By observing at frequent time intervals that no photons have leaked from the
cavity a conditional evolution is constructed. Emission of a photon
corresponds to a quantum jump \cite{barchielli} and the evolution is
described here within this framework (quantum jump approach
\cite{hegerfeldt}), where the system evolves according to a non-Hermitian
Hamiltonian due to its coupling with the environment. The combination of a
strong $\kappa$ and the detector forces the system to remain in the DFS$_c$ by a
mechanism called {\it environment-induced quantum Zeno effect}
\cite{Misra,beige2}. Combined with the adiabatic procedure described above, it
has the result that weak laser couplings between the ground and excited
atomic levels do not move the system out of the initially populated
DFS$_c$. In other words no population of the cavity occurs for a long
time interval.

In particular, we shall apply {\it common} laser addressing to the two atoms
with a possible phase or amplitude difference in their Rabi
frequencies. Consider a laser tuned between an auxiliary ground state
$|\sigma\rangle$ and $|2\rangle$ and a second one tuned between $|1\rangle$
and $|2\rangle$. The conditional Hamiltonian that describes the evolution of
the system is given by
\bq
H_{\text{cond}}= && g \hbar \sum_{i=1}^2 \big(|2\ran_i \lan 1| b + \text{h.c.}\big)
\no \no
&&
+{1 \over 2} \hbar \big( \Omega_1|\sigma\ran_1 \lan 2| + \Omega_1^\prime
|\sigma\ran_2 \lan 2| 
\no \no
&&
\Sp \Sp + \Omega_2|1\ran_1 \lan 2| + \Omega_2^\prime |1\ran_2 \lan 2| + \text{h.c.} \big)
\no \no
&&
-{i \kappa \over 2} \hbar b^\dagger b +{\Delta-i \Gamma \over 2} \hbar 
\sum_{i=1}^2 |2 \ran_i \lan 2| \,\, ,
\label{condham}
\eq
where $b$ is the annihilation operator of the cavity mode, $\Delta$ is the
cavity and laser detuning and the subscript $i$ on the states denotes
different atoms. For different values of the atomic spontaneous emission rate,
$\Gamma$, we may reconfigure the detuning $\Delta$ and amplitudes $\Omega_i$
in order to optimize the fidelity of the gates and their success rate.

Consider initially the effect of $\kappa$ and small $\Omega$'s on the
evolution of the system. Similarly to Beige {\it et al.}
\cite{beige,beige1}, $\Omega_i$ is such that it performs a transition
slow enough to keep the state of the system inside the DFS$_c$, i.e.
\be
|R| \ll {g^2 \over \kappa} \Sp \text{and} \Sp \kappa \,\, ,
\label{cond}
\ee
where $R$ is the state transition rate from the DFS$_c$. For this
configuration 
the effective Hamiltonian (see \cite{beige2}) is given by $H_{\text{eff}}=
{\bf P}_{\text{DFS}_c} H_{\text{laser}} {\bf P}_{\text{DFS}_c}$, where ${\bf
P}_{\text{DFS}_c}$ is the projector in the cavity decoherence-free
space, while 
$H_{\text{laser}}$ is the laser part of Hamiltonian (\ref{condham}). With the
laser amplitudes tuned as $\Omega_1=\Omega_1^\prime=\sqrt{2} \Omega$,
$\Omega_2 -\Omega_2^\prime=2 \bar \Omega$ and on the basis $|A\rangle \equiv
(|\sigma 1 \rangle -|1 \sigma \rangle)/\sqrt{2}$, $|11\rangle$ and
$|\alpha\rangle$ the effective Hamiltonian becomes
\be 
H_{\text{eff}}=\hbar {\Delta \over 2} |\alpha\ran \lan \alpha|+ \hbar \left(\Omega
|\alpha \rangle \langle A| + \bar \Omega |\alpha \rangle \langle 11|
+\text{h.c.} \right) \,\, .
\nonumber
\ee 
Note that the states $|\alpha\ran$ and $|A\rangle$ do not belong in the
two-qubit computational space spanned by $|ij\rangle$ for $i,j=0,1$. Two of
the eigenvalues of this Hamiltonian, $E_{1,2}=\Delta/4 \pm \sqrt{|\Omega|^2+
|\bar \Omega|^2+(\Delta/4)^2}$, have eigenvectors that occupy the
antisymmetric state $|\alpha \rangle$, while the third eigenvalue, $E_3=0$,
corresponds to the eigenvector $(-\bar \Omega, \Omega, 0)/\sqrt{|\Omega|^2+
|\bar \Omega|^2}$. The latter is the only one with zero component on
the $|\alpha\rangle$ 
state and hence on the excited state $|2\rangle$. As a consequence, adiabatic
transfer of population can occur between states $|11\rangle$ and $|A\rangle$
by slowly varying the laser amplitudes $\Omega$ and $\bar \Omega$ in such a
way that the population remains on the third eigenstate without ever
populating the decaying level $2$. 
\begin{center}
\begin{figure}[ht]
\centerline{
\put(265,287){$P_0$}
\put(235,255){$|A\ran$}
\put(160,260){$|11\ran$}
 \epsffile{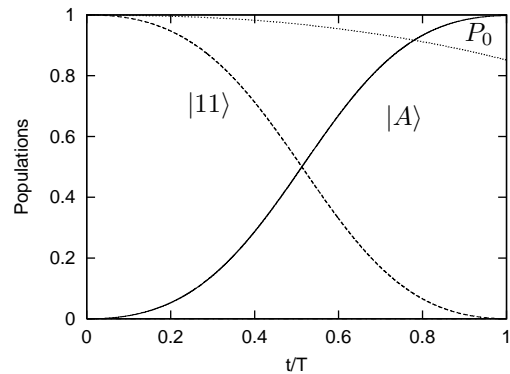}
}
\vspace{-5.5cm}
 \caption[contour]{\label{p001}
Adiabatic evolution for $\Gamma=0.1g$ and $\kappa=0.1g$. The
population of the state 
$|11\ran$ is completely transferred to the state $|A\ran$ with success rate
$P_0=0.852$ and fidelity $F=0.999$. For this transition the overall time of
the adiabatic procedure, which is performed with linear ramp for the laser
pulse, is $T=5 \cdot 10^4/g$.
}
\vspace{-0.5cm}
\end{figure}
\end{center}

This adiabatic passage employed to eliminate the excited level $2$ is
enhanced by the Zeno effect \cite{beige2} applied to the combination
of strong $\Gamma$ and detectors that observe emitted photons from the
atoms. The projection to the atomic {\it and} cavity decoherence 
free subspace has as a result that an initial population of the
eigenstate with $E_3=0$ remains there though out the adiabatic
procedure while the two other eigenstates are projected out. 
Note that if state $|\si\ran$ was not
employed, but instead we used state $|0\ran$, then the state $|A\ran$ could be
populated initially and the transfer described above would be impossible. For
amplitudes satisfying $|\Omega|/|\bar \Omega|=\tan(\theta/2)$ and with phase
difference $\phi= \phi_1-\phi_2$ the eigenstate corresponding to the zero
eigenvalue takes the form of the dark state of the system $|D\rangle =\cos
{\theta \over 2} |11\rangle - \sin{\theta \over 2} e^{i \phi} |A\rangle$. In
Fig. \ref{p001} the $\theta$ transition from zero to $\pi$ is depicted where
$\Omega_{\text{max}}/\sqrt{2}=\bar\Omega_{\text{max}}=0.018 g$, $\Gamma=
\kappa=0.1 g$ and $\Delta= 0.02 g$. The simulation is performed with the
evolution dictated by the conditional Hamiltonian (\ref{condham}).
The probability of no photon emission from the cavity, or success rate, is
$P_0=0.852$, while the fidelity is $F=0.999$. Along this evolution the
antisymmetric state $|\alpha\ran$ does not get populated. Such a procedure
resembles the STIRAP process, which produces population transfer between
ground states of one atom, but now the transfer is between states of two
atoms.

It is possible to optimize the fidelity, $F$, and the probability for no
photon emission, $P_0$, of this transition with
respect to different values of the detuning $\Delta$ and the maximum
Rabi frequency $\Omega_{\text{max}}$. In the following simulations we
vary both 
$\Delta$ and $\Omega_{\text{max}}$ for values of the spontaneous
atomic emission $\Gamma=0.1g$ and leakage of the cavity
$\kappa=0.1g$, that are within the relatively strong coupling regime
$g^2  > \kappa \Gamma$. 
\begin{center}
\begin{figure}[ht]
\centerline{
\epsffile{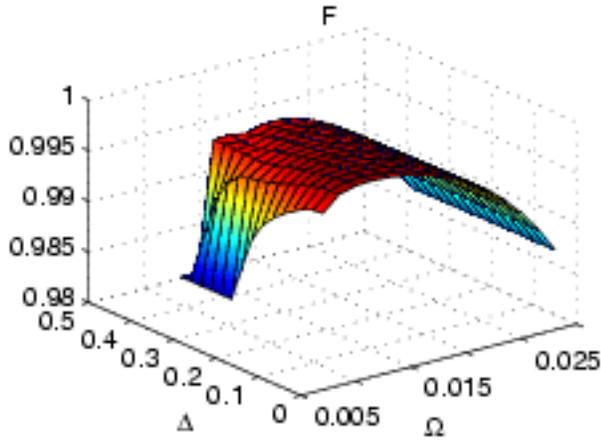}
}
\vspace{0.2cm}
 \caption[contour]{\label{double}
The fidelity for the adiabatic transfer of population for
$\Gamma=0.1g$ and $k=0.1g$. We observe maximum at $\Omega=0.0145g$ and
$\Delta\approx0.024g$ with fidelity $F=0.999$.
}
\end{figure}
\end{center}
\vspace{-1cm}
\begin{center}
\begin{figure}[ht]
\centerline{
 \epsffile{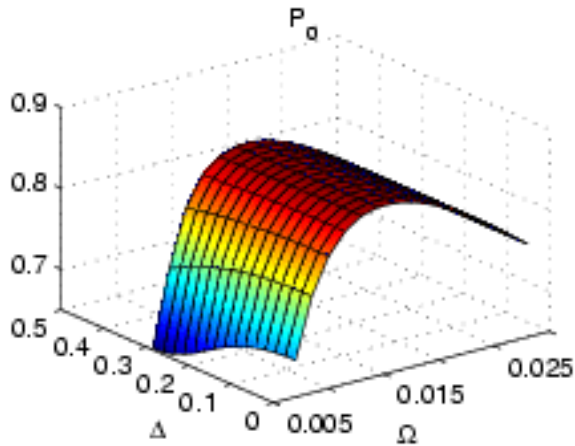}
}
\vspace{0.2cm}
 \caption[contour]{\label{double1}
The probability for no photon emissions for the adiabatic transfer 
with $\Gamma=0.1g$ and $k=0.1 g$. We observe maximum for
$\Omega=0.0145g$ and any $\Delta$ with $P_0=0.858$.
This coincides with the range the fidelity is also maximum.
}
\vspace{-0.8cm}
\end{figure}
\end{center}
Thus we obtain Fig. \ref{double} where the maximum fidelity is given
for $\Omega=0.0145g$ and $\Delta$ less than $0.024g$. For these ranges of
$\Delta$ and $\Omega_{\text{max}}$ the corresponding value of the
probability for no photon emission is given in Fig. \ref{double1}
which is within acceptable limits. With a close inspection we observe
that the maximum success rate increases when $\Delta$ takes a non-zero
but small value. For large values of the detuning the fidelity drops
down as the eigenvalues of $H_{\text{eff}}$ come closer and the
adiabatic procedure fails. For strong lasers the success rate improves
slightly for larger $\Delta$ as expected, but the fidelity decreases
as we move out of the regime (\ref{cond}) of weak lasers where the
projection in the DFS$_c$ holds.

There is the possibility of realizing dynamical and geometrical gates
between the two atoms. Consider the following evolution. Take $\theta$
from zero to $\pi$, then apply a $2 \pi$ pulse to transform the state
$|\sigma\ran$ to $-|\sigma\ran$ for both atoms and then take $\theta$
back to zero. Then the
state $|11\ran$ acquires an overall minus sign, while the rest of the
computational states $|00\rangle$, $|01\rangle$ and $|10\rangle$ remain
unchanged. This is a conditional phase shift,
$\text{diag}(1,1,1,e^{i\varphi})$ with $\varphi=\pi$. As an additional
application holonomic gates \cite{pachos1} can be constructed in the same
fashion as in the ion trap model proposed by Duan {\it et al.} \cite{Duan}.
By continuously changing the variables $\theta$ and $\phi$ starting from
$\theta=0$ one can perform a cyclic adiabatic evolution on the $(\theta,
\phi)$ plane described by a loop $C$. At the end of this evolution the state
$|11\rangle$ acquires a Berry phase and the overall gate is given by the
holonomy $\Gamma(C)= \exp ( i |11\rangle\langle 11| \,\,
\varphi_{\text{Berry}} )$, where $\varphi_{\text{Berry}}= \int \sin \theta
\,\, d \theta d \phi$ and the integration runs over the surface the loop $C$
encloses. The evolution has the form of a conditional phase-shift gate. This
model has the experimental advantages presented in \cite{Duan} and in addition
{\it no} cooling of the trapped ions' modes is required beyond Doppler cooling.

\begin{center}
\begin{figure}[ht]
\centerline{
\put(-20,143){$P_{{3 \over 2}}$}
\put(-20,110){$P_{{1 \over 2}}$}
\put(-20,55){$D_{{3 \over 2}}$}
\put(-20,25){$S_{{1 \over 2}}$}
\put(25,125){$2\pi$}
\put(45,80){$\sigma_+$}
\put(65,80){$\sigma_-$}
\put(95,80){$g$}
\put(10,43){$|\sigma\ran$}
\put(53,15){$|0\ran$}
\put(103,15){$|1\ran$}
\put(70,-10){1-qubit}
 \epsffile{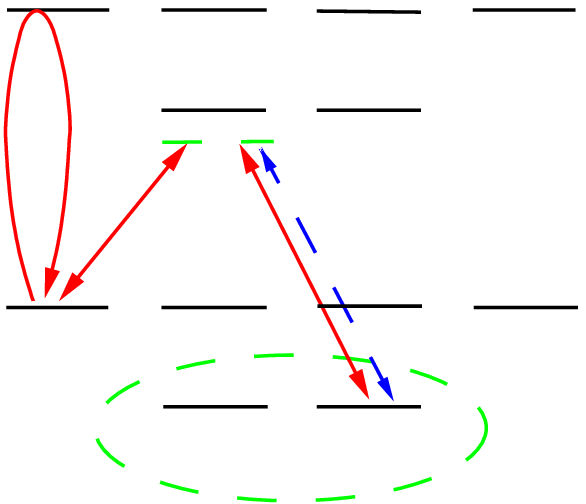}
}
\vspace{0.5cm}
 \caption[contour]{\label{calcium}
Application of the proposed scheme to the Calcium ion Ca$^+$.
}
\vspace{-0.5cm}
\end{figure}
\end{center}

This proposal can be implemented within the ion trap and cavity
experiments performed, e.g. in Garching \cite{wolfgang}. There, trapped Calcium
ions, Ca$^+$, are placed inside a cavity and can be addressed with laser
fields. The application of our proposal to the internal levels of Ca$^+$ is
presented in Fig. \ref{calcium}.
The states $|0\ran$ and $|1\ran$ are the ones which encode the qubit. 
Level $\si$ is the auxiliary level that enables to perform conditional 
transfer of population without affecting the qubit state $|0\ran$. By
the end of the manipulation it is emptied.

In this proposal we describe two-qubit conditional phase-shift gates for ions
trapped in a cavity. Together with one-qubit rotations they consist of a
universal set of gates. The cavity and the atomic spontaneous emission rates
are considered here to be close to the atom-cavity coupling. An
adiabatic transition between states of the two atoms is performed in such a
way that the cavity mode and the excited atomic levels are virtually populated
thus avoiding the problem of their decoherence. This population transfer
resembles the STIRAP procedure, hence it enjoys the experimental advantage of
the final state being independent of the exact intermediate value of the
amplitude of the laser beams. By observing at frequent time intervals
for emitted photons from the atom and the cavity the adiabatic
transition is assisted by the Zeno effect guarantying that the
population will remain in the decoherence free subspace by projecting
out the excited atomic and cavity states. It has been shown here that
the success rate
and the fidelity of the gates are close to unity, allowing construction of a
system for quantum computation in the presence of decoherence. Gates are
constructed dynamically as well as geometrically in order to take advantage of
the additional fault-tolerant features of geometrical quantum computation
\cite{ellinas}.

{\em Acknowledgements.} A. Beige, I. Cirac, W. Lange and M. Plenio are thanked for
discussions. This work was partially supported by the European TMR network for 
Quantum Information.


\end{multicols}

\begin{references}

\bibitem{pellizzari} T. Pellizzari, S. Gardiner, J. Cirac, P. Zoller,
 Phys. Rev. Lett. {\bf 75}, 3788 (1995).

\bibitem{dfs} G. Palma, K.-A. Suominen, A. Ekert,
 Proc. Roy. Soc. London Ser. A {\bf 452}, 567 (1996); P. Zanardi,
 M. Rasetti, Phys. Rev. Lett. {\bf 79}, 3306 (1997); D. Lidar, I. Chuang,
 K. Whaley, Phys. Rev. Lett. {\bf 81}, 2594 (1998); L. Duan, G. Guo,
 Phys. Rev. A {\bf 58}, 3491 (1998).

\bibitem{molmer} K. M{\o}lmer, A. S{\o}rensen, Phys. Rev. Lett. {\bf 82},
 1835 (1999); A. S{\o}rensen, K. M{\o}lmer, Phys. Rev. A {\bf 62}, 022311
 (2000). 

\bibitem{beige} A. Beige, D. Braun, B. Tregenna, P. Knight, Phys. Rev. Let.
 {\bf 85}, 1762 (2000).

\bibitem{Duan1} L.-M. Duan, G. C. Guo, Phys. Rev. A {\bf 58}, 3491 (1998).

\bibitem{Zanardi} P. Zanardi, F. Rossi, Phys. Rev. B {\bf 59}, 8170 (1999).

\bibitem{Bacon} D. A. Lidar, D. Bacon, J. Kempe, K.B. Whaley, Phys. Rev. A
 {\bf 63}, 022306 (2001); J. Kempe, D. Bacon, D. A. Lidar, K. B. Whaley,
 quant-ph/0004064. 

\bibitem{plenio} M. Plenio, S. Huelga, A. Beige, P. Knight, Phys. Rev. A
 {\bf 59}, 2468 (1999).

\bibitem{haroche} S. Osnaghi, P. Bertet, A. Auffeves, P. Maioli, M. Brune,
 J.M. Raimond, S. Haroche, ``Coherent control of an atomic collision in a
 cavity", quant-ph/0105063.

\bibitem{solano} E. Solano, R.L. de Matos Filho, N. Zagury, Phys. Rev. A
 {\bf 59}, R2539 (1999); {\bf 61}, 029903(E) (2000).

\bibitem{comment} This is not a necessary condition; it merely simplifies the
 presentation. 

\bibitem{barchielli} A. Barchielli, V. Belavkin, J. Phys. A {\bf 24}, 1495
 (1991). 

\bibitem{hegerfeldt} G. Hegerfeldt, D. Sondermann, Quantum and
 Semiclassical Opt. {\bf 8}, 121 (1996).

\bibitem{Misra} B. Mirsa, E. C. G. Sudarshan, J. Math. Phys. {\bf 18}, 756
 (1977). 

\bibitem{beige2} A. Beige, D. Braun, P. L. Knight, New J. Phys. 2, 22 (2000).

\bibitem{beige1} B. Tregenna, A. Beige, P. Knight, ``Quantum Computing in the
 Dark", quant-ph/0109006.

\bibitem{pachos1} P. Zanardi, M. Rasetti, Phys. Lett. A264 (1999) 94; 
 J. Pachos, ``Quantum Computation by Geometrical Means",
 published in the AMS Contemporary Math Series 
 volume entitled ``Quantum Computation \& Quantum Information Science".

\bibitem{Duan} L.-M. Duan, J. I. Cirac, P. Zoller, Science 2001 June 1;
 292: 1695.

\bibitem{berman} ``Cavity quantum electrodynamics", Edited by P. Berman, Academic
 Press (1994).

\bibitem{wolfgang} G. R. Guth\"ohrlein, M. Keller, K. Hayasaka, W. Lange, 
 H. Walther, ``A single ion as a nanoscopic probe of an optical field'',
 Nature, in print. 

\bibitem{ellinas} D. Ellinas, J. Pachos, Phys. Rev. A {\bf 64}, 022310 (2001).

\end{references}
\end{document}